\documentclass[prl,aps,superscriptaddress,showpacs,twocolumn,notitlepage]{revtex4-1}
\usepackage{bm}
\usepackage{amsfonts}
\usepackage[dvips]{graphicx}
\usepackage{mathrsfs}
\usepackage[intlimits]{amsmath}
\usepackage[colorlinks, citecolor=red]{hyperref}

\hypersetup{colorlinks,
linkcolor=blue,          citecolor=blue,        filecolor=blue,      urlcolor=blue           }
\graphicspath{{../figures/}}

\begin{document}

\title{Hamiltonian tomography for quantum many-body systems with arbitrary couplings}

\author{Sheng-Tao Wang}
\affiliation{Department of Physics, University of Michigan, Ann Arbor, Michigan 48109, USA}
\author{Dong-Ling Deng}
\affiliation{Department of Physics, University of Michigan, Ann Arbor, Michigan 48109, USA}
\affiliation{Condensed Matter Theory Center and Joint Quantum Institute, Department of Physics, University of Maryland, College Park, MD 20742, USA}
\author{L.-M. Duan}
\affiliation{Department of Physics, University of Michigan, Ann Arbor, Michigan 48109, USA}

\date{\today }

\begin{abstract}
Characterization of qubit couplings in many-body quantum systems is essential for benchmarking quantum computation and simulation. We propose a tomographic measurement scheme to determine all the coupling terms in a general many-body Hamiltonian with arbitrary long-range interactions, provided the energy density of the Hamiltonian remains finite. Different from quantum process tomography, our scheme is fully scalable with the number of qubits as the required rounds of measurements increase only linearly with the number of coupling terms in the Hamiltonian. The scheme makes use of synchronized dynamical decoupling pulses to simplify the many-body dynamics so that the unknown parameters in the Hamiltonian can be retrieved one by one. We simulate the performance of the scheme under the influence of various pulse errors and show that it is robust to typical noise and experimental imperfections.
\end{abstract}

\maketitle

\textit{Introduction.---}Physicists have been striving to understand and
harness the power of quantumness since the establishment of the quantum
theory. With the flourishing of quantum information science in recent
decades \cite{nielsen2010quantum, Ladd2010Quantum}, numerous
breakthroughs---both in theory and in experiment---helped to frame a clearer
goal: it is the entanglement and the exponentially growing Hilbert space
that distinguishes quantum many-body systems from classical systems \cite%
{Feynman1982Simulating, Horodecki2009Entanglement, Cirac2012Goals}. To fully
leverage the quantum supremacy, a vital step is to verify and benchmark the
quantum device. The standard techniques of quantum state and process
tomography \cite{Vogel1989Determination, Smithey1993Measurement,
Chuang1997Prescription, James2001Measurement, Haffner2005Scalable,
Lvovsky2009Continuous}, however, are plagued by the same exponential growth
of dimensions \cite{paris2004quantum}. A related problem is to directly
identify Hamiltonians, the generators of quantum dynamics. They can often be
specified by fewer number of parameters that scales polynomially with the
system size.

Hamiltonian tomography for generic many-body systems is nevertheless a
daunting task. The way to extract information of unknown parameters in a
Hamiltonian is by measuring certain features of its generated dynamics. To
make this possible, one has to solve the dynamics generated by the
Hamiltonian to make a definite connection between its dynamical features and
the Hamiltonian parameters. However, for general many-body Hamiltonians,
their dynamics are extremely complicated and intractable by numerical
simulation as the simulation time increases exponentially with the size of
the system. Progress in this direction has mostly be on small systems \cite{Schirmer2004Experimental, Cole2005Identifying, Cole2006Precision, Devitt2006Scheme, Senko2014Coherent} or special many-body systems which are
either exactly solvable due to many conserved operators, of limited
Hilbert space dimensions amenable to numerical simulation, or short-range interacting systems \cite
{Burgarth2009Indirect, Burgarth2009Coupling, Burgarth2011Indirect,
DiFranco2009Hamiltonian, Zhang2014Quantum, daSilva2011Practical, Wiebe2014Hamiltonian}.

In this paper, we propose a scheme to achieve Hamiltonian tomography for
general many-body Hamiltonians with arbitrary long-range couplings between
the qubits. The key idea is to simplify the dynamics generated by a general
many-body Hamiltonian through application of a sequence of dynamical
decoupling pulses on individual qubits. Dynamical decoupling (DD) is a
powerful technique that uses periodic fast pulses to suppress noise and
average out unwanted couplings between the system and the environment \cite{Gullion1969New, Viola1998Dynamical, Viola1999Dynamical, Duan1999Suppressing,
Zanardi1999Symmetrizing, Khodjasteh2005Fault, Khodjasteh2007Performance,
Uhrig2007Keeping, Yang2011Preserving, Souza2011Robust,
Souza2012Robust, Alvarez2010Perfect, Morton2006Bang, Biercuk2009Optimized, Du2009Preserving, West2010High}. We apply a sequence of synchronized DD pulses on a pair of
qubits, which forms a small target system that has coupling with the rest of the
qubits in the many-body Hamiltonian, the effective environment. The DD
pulses keep the desired couplings within this target system intact while
average out its couplings with all the environment qubits. The dynamics
under the DD pulses become exactly solvable, from which we can perform a
tomographic measurement to determine the coupling parameters within this
small target system   \cite{Schirmer2004Experimental, Cole2005Identifying,
Cole2006Precision, Devitt2006Scheme}. We then scan the DD
pulses to different pairs of qubits to measure all the other coupling terms
in the Hamiltonian. We assume the ability to address individual qubits,
which is realistic for many experimental platforms, such as trapped ions \cite%
{Schmidt-Kaler2003Realization, Johanning2009Individual, Crain2014individual}%
, cold atoms \cite{Sherson2010single, Bakr2010probing, Weitenberg2011Nature}%
, and solid-state qubit systems \cite{Devoret2013Science,  Barends2014Superconducting, Barends2015Digital, Salathe2015Digital}. Several features make the scheme amenable to
experimental implementation. First of all, applying the DD pulse sequence is
a standard procedure in many experiments. Post-processing of data is
straightforward as it only requires one or two parameter curve fitting. In
addition, we demonstrate with explicit numerical simulation that the scheme
is robust to various sources of errors in practical implementation, such as
the remnant DD coupling error, measurement uncertainties, and different types
of pulse errors.

\begin{figure}[t]
\includegraphics[trim=0cm 0cm 0cm 0cm, clip,width=0.48\textwidth]{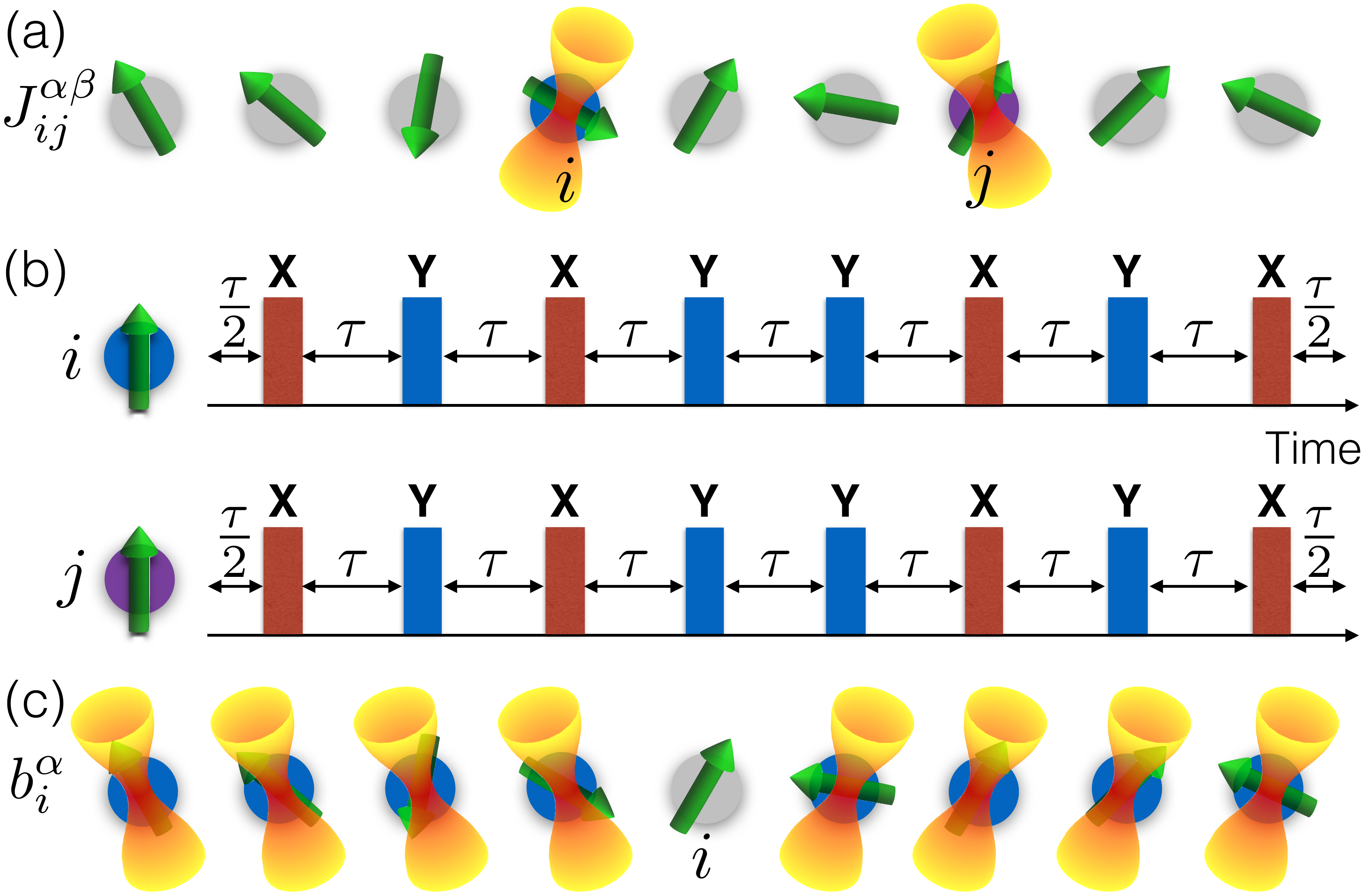}
\caption{(color online). Schematics for the tomography procedure. (a) To map
out the coupling coefficients $J_{ij}^{\protect\alpha \protect\beta}$, a
synchronized DD sequence is applied to spins $i$ and $j$. Both spins will be
decoupled from the rest of the system. (b) The $XY$-$8$ DD sequence on spins
$i$ and $j$ to probe the parameters of the Hamiltonian in Eq.~\eqref{Eq:Ham2}%
. The initial state is for instance prepared to the $| 00 \rangle$ state for
the two spins. (c) To retrieve information about the local fields $b_{i}^{%
\protect\alpha}$, $XY$-$8$ pulse sequences are applied to the environment
spins to decouple spin $i$ from the rest.}
\label{Fig:Schematics}
\end{figure}

\textit{Scheme for Hamiltonian tomography.---}The system we have in mind is
the most general Hamiltonian with two-body qubit interactions
\begin{equation}
H=\sum_{\alpha ,\beta ,m<n}J_{mn}^{\alpha \beta }\sigma _{m}^{\alpha }\sigma
_{n}^{\beta }+\sum_{m,\alpha }b_{m}^{\alpha }\sigma _{m}^{\alpha },
\label{Eq:HamGeneral}
\end{equation}%
where $J_{mn}^{\alpha \beta }$ characterizes the coupling strength between
spins $m$ and $n$ for the $\alpha ,\beta $ components, and $b_{m}^{\alpha }$
represents the local field on spin $m$; $\sigma ^{\alpha }\left( \sigma
^{\beta }\right) $ are the Pauli matrices along the $\alpha $ $(\beta )$
direction with $\alpha ,\beta \in (x,y,z)$. To adopt consistent notations
throughout the text, we use $m,n$ to denote a general spin label and $i,j$
to refer to the specific  \textit{target spins} that we are probing with the DD
pulses, calling the rest of the spins as \textit{environment spins}. The terms spin and qubit are used interchangeably. 
Let the
energy unit of the Hamiltonian be $J$, chosen to be the largest magnitude of
all coefficients, so $J_{mn}^{\alpha \beta }/J$ and $b_{m}^{\alpha }/J$ are
bounded between $-1$ and $1$. In order to map out the coupling coefficient $%
J_{ij}^{\alpha \beta }$ for the target spins, we propose to decouple
these two spins from the environment spins by a synchronized DD pulse
sequence. A synchronized $XY$-$4$ sequence applied to both spins will
average out their interactions with other spins while preserving the
two-spin coherence (see Fig.~\ref{Fig:Schematics}(a-b) for the schematic and
the pulse sequence). Basically, only those interactions that commute with
the DD sequence will survive. More rigorously, the evolution operator in one
period is
\begin{align}
U_{1}& =U_{0}^{1/2}\sigma _{i}^{x}\sigma _{j}^{x}U_{0}\sigma _{i}^{y}\sigma
_{j}^{y}U_{0}\sigma _{i}^{x}\sigma _{j}^{x}U_{0}\sigma _{i}^{y}\sigma
_{j}^{y}U_{0}^{1/2}  \notag \\
& =e^{-i4\tau (J_{ij}^{xx}\sigma _{i}^{x}\sigma _{j}^{x}+J_{ij}^{yy}\sigma
_{i}^{y}\sigma _{j}^{y}+J_{ij}^{zz}\sigma _{i}^{z}\sigma
_{j}^{z}+B)+O(J^{2}\tau ^{2})},
\end{align}%
where $U_{0}=e^{-iH\tau }$, $\tau $ is the time interval between two
consecutive pulses, and $B$, the bath, includes all terms of the Hamiltonian
that only acts on environment spins. See Supplemental Material for the
detailed derivation \cite{HamTom:Sup}. To bound the error term to $%
O(J^{2}\tau ^{2})$, we assume $\sum_{n}J_{in}^{\alpha \beta }=O(J)$, i.e.,
the interaction strength decays rapidly with spin separation distance so that the energy density of the Hamiltonian is bounded by a constant. This condition is satisfied for any finite systems as in the experiment with arbitrary interactions. 
In the thermodynamic limit, it is also a reasonable assumption for any physical systems whose energy is extensive. It may also be related to the generalized Lieb-Robinson bound for systems with long-range interactions \cite{Hastings2006Spectral, Nachtergaele2006Lieb, Gong2014Persistence, FossFeig2015Nearly}. 
The $XY$-$8$ pulse sequence, which is the concatenation of $XY$-$4$ sequence
with its time-reversal, eliminates the error term to the third order $%
O(J^{3}\tau ^{3})$. Fig.~\ref{Fig:Schematics}(b) shows the $XY$-$8$ DD pulse
sequence. Hence, in the Hilbert subspace of the target spins, the effective Hamiltonian is
\begin{equation}  \label{Eq:Ham2} 
H_{2\text{-spin}}=c_{1}\sigma _{i}^{x}\sigma _{j}^{x}+c_{2}\sigma
_{i}^{y}\sigma _{j}^{y}+c_{3}\sigma _{i}^{z}\sigma _{j}^{z},
\end{equation}%
where we use $c_{1}\equiv J_{ij}^{xx},c_{2}\equiv J_{ij}^{yy},c_{3}\equiv
J_{ij}^{zz}$ to simplify the notation. The effective two-spin unitary
evolution after $N_{c}$ cycles of $XY$-$8$ sequence is
\begin{align*}
& U_{2\text{-spin}}= \\
& \!\!\left(
\begin{array}{cccc}
\!\tfrac{\cos \left( (c_{1}-c_{2})T\right) }{e^{ic_{3}T}} & \!0 & \!0 & \!%
\tfrac{\sin \left( (c_{1}-c_{2})T\right) }{ie^{ic_{3}T}} \\
\!0 & \!\tfrac{\cos \left( (c_{1}+c_{2})T\right) }{e^{-ic_{3}T}} & \!\tfrac{%
\sin \left( (c_{1}+c_{2})T\right) }{ie^{-ic_{3}T}} & \!0\vspace{0.1cm} \\
\!0 & \!\tfrac{\sin \left( (c_{1}+c_{2})T\right) }{ie^{-ic_{3}T}} & \!\tfrac{%
\cos \left( (c_{1}+c_{2})T\right) }{e^{-ic_{3}T}} & \!0 \\
\!\tfrac{\sin \left( (c_{1}-c_{2})T\right) }{ie^{ic_{3}T}} & \!0 & \!0 & \!%
\tfrac{\cos \left( (c_{1}-c_{2})T\right) }{e^{ic_{3}T}}%
\end{array}%
\!\right) ,
\end{align*}%
where $T=8N_{c}\tau $ is the total time. From the above expression, one may
notice that the Hamiltonian parameters can be retrieved by preparing a
particular initial state and measuring its time-evolved output probability
in a given basis. In particular, we have
\begin{align*}
P_{|+\text{I}\rangle \rightarrow |00\rangle }& =\left\vert \langle 00|U_{2%
\text{-spin}}|\!+\!\text{I}\rangle \right\vert ^{2}=\tfrac{1}{4}\left[
1+\sin (2(c_{1}-c_{2})T)\right]  \\
P_{|+\text{I}\rangle \rightarrow |10\rangle }& =\left\vert \langle 10|U_{2%
\text{-spin}}|\!+\!\text{I}\rangle \right\vert ^{2}=\tfrac{1}{4}\left[
1+\sin (2(c_{1}+c_{2})T)\right]  \\
P_{|0\text{I}\rangle \rightarrow |++\rangle }& =\left\vert \langle
+\!+\!|U_{2\text{-spin}}|0\text{I}\rangle \right\vert ^{2}\!=\!\tfrac{1}{4}%
\left[ 1+\sin (2(c_{2}-c_{3})T)\right]
\end{align*}%
where $|+\rangle =\frac{1}{\sqrt{2}}(|0\rangle +|1\rangle )$ and $|\text{I}%
\rangle =\frac{1}{\sqrt{2}}(|0\rangle +i|1\rangle )$ are the rotated basis.
The coupling strengths $c_{1},c_{2}$ and $c_{3}$ can be extracted from the
oscillation frequencies of these three sets of measurements at various time
points. These particular sets are not the only suite to extract those
parameters. They are chosen for the convenience in fitting and in state
preparation. Only product states of the two target spins, disentangled from the
rest, are required. We also remark that the error incurred is $%
O(N_{c}J^{3}\tau ^{3})$, so one needs $J\tau \ll 1$ for a robust decoupling
scheme. In a similar fashion, one can retrieve all other coupling
coefficients. Let us denote the synchronized $XY$-$8$ DD pulse sequence as $%
X_{i}X_{j}$-$Y_{i}Y_{j}$-$8$ to show explicitly the particular pulses on
specific spins. Replacing the sequence with $X_{i}Y_{j}$-$Y_{i}Z_{j}$-8 ($%
Y_{i}X_{j}$-$Z_{i}Y_{j}$-8) pulses, we will be able to extract the
coefficients $J_{ij}^{xy},J_{ij}^{yz}$ and $J_{ij}^{zx}$ ($%
J_{ij}^{yx},J_{ij}^{xz}$ and $J_{ij}^{zy}$), respectively.

\begin{figure*}[t]
\includegraphics[trim=0cm 0cm 0cm 0cm, clip,width=\textwidth]{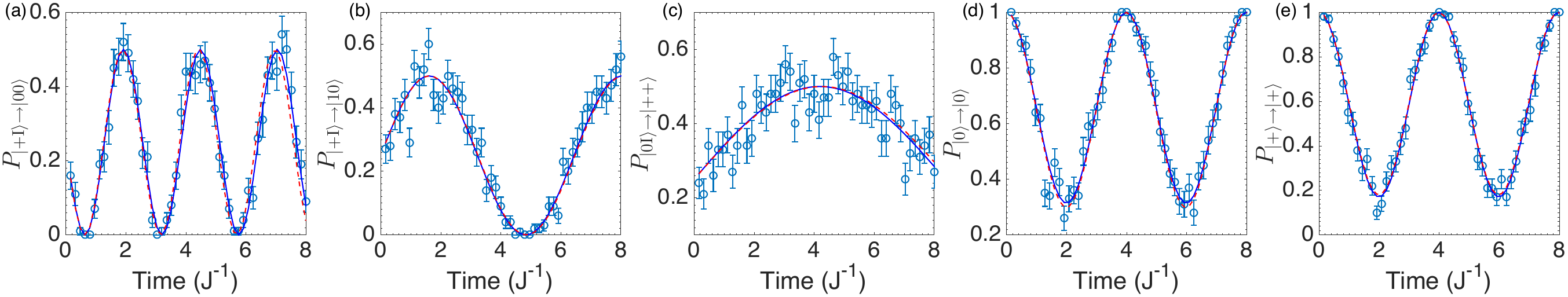}
\caption{(color online). Numerical simulation and curving fitting results.
(a)-(c) are used to retrieve $J_{79}^{xx}$, $J_{79}^{yy}$ and $J_{79}^{zz}$
between spins 7 and 9. (d) and (e) are used to extract $b_{6}^{x}$, $%
b_{6}^{y}$ and $b_{6}^{z}$ for spin 6. Each measurement data $p_{m}$ are
drawn from the binomial distribution with the true probability $p$ as the
mean and $p(1-p)/N_{m}$ as the variance. The measurement uncertainty of each
point is thus $\protect\sqrt{ p_{m}(1-p_{m})/N_{m} }$. The blue solid lines
are the best-fit lines with the simulated experimental data $p_{m}$, and red
dashed lines are the theoretical ones generated by the true Hamiltonian
parameters. Pulse errors are not included in these plots, so any
discrepancies stem from the remnant DD coupling error and measurement
uncertainties. Other parameters used are $N=12$, $\protect\tau J=0.01$, $%
N_{m}=100$, $N_{t}=50$.}
\label{Fig:Estimation}
\end{figure*}

By scanning the DD pulses to different target pairs, the above procedure recovers all the coupling coefficients $J_{mn}^{\alpha
\beta }$. The retrieval of local field coefficients follows a similar
approach. We now need to decouple the particular spin $i$ from the rest
without contaminating its own spin term $b_{i}^{\alpha }\sigma _{i}^{\alpha }
$. Shining a $XY$-$8$ DD sequence on spin $i$ removes all information about $%
b_{i}^{\alpha }$ too. Instead, one could address all the environment spins
with $XY$-$8$ pulses, and decouple them from spin $i$ (alternative schemes
are discussed in the Supplemental Material \cite{HamTom:Sup}). This scheme
will be very robust to pulse errors, since no laser pulses are directly
applied to the target spin [Fig.~\ref{Fig:Schematics}(c)]. The effective
single-spin Hamiltonian is thus $H_{1\text{-spin}}=b_{i}^{x}\sigma
_{i}^{x}+b_{i}^{y}\sigma _{i}^{y}+b_{i}^{z}\sigma _{i}^{z}$ with a unitary
evolution $U_{1\text{-spin}}=e^{-iH_{1\text{-spin}}T}$ executing a spin
rotation on the Bloch sphere. Again, by preparing a particular state and
measuring its time evolution, we get $P_{|0\rangle \rightarrow |0\rangle }=1+%
\left[ \left( b_{i}^{z}/b\right) ^{2}-1\right] \sin ^{2}(bT)$ and $%
P_{|+\rangle \rightarrow |+\rangle }=1+\left[ \left( b_{i}^{x}/b\right)
^{2}-1\right] \sin ^{2}(bT)$, where $b=\sqrt{%
(b_{i}^{x})^{2}+(b_{i}^{y})^{2}+(b_{i}^{z})^{2}}$ is the magnitude of the
Bloch vector. These two sets of measurements will determine $%
b_{i}^{x},b_{i}^{y}$ and $b_{i}^{z}$ up to a sign. The correct signs from
the remaining discrete set can be picked out by measuring $P_{|+\rangle
\rightarrow |0\rangle }$ and $P_{|\text{I}\rangle \rightarrow |0\rangle }$
at a single time point \cite{HamTom:Sup}.

The complete scheme applies to any generic Hamiltonian with interacting
qubits. In the most general case, one needs to determine $9N(N-1)/2+3N$
coefficients. However, in many physical systems, the particular form of the
interaction is known and/or the interaction often decays fast enough that
one can significantly reduce the number of measurements required. In
particular, if $J_{mn}^{\alpha \beta }$ can be truncated at some spin
separation distance in the case of short-range interactions, the number of
measurements will be linear with the system size $N$. In the following, we
numerically simulate the experimental procedure for the most general
Hamiltonian, taking into account various sources of errors, including the
remnant DD coupling error, measurement uncertainties, and different forms of
pulse errors.

\textit{Numerical simulation.---}We consider the general Hamiltonian given
in Eq.~\eqref{Eq:HamGeneral} with coefficients $J_{ij}^{\alpha \beta }/J$
and $b_{i}^{\alpha }/J$ randomly drawn from $-1$ to $1$. In our
finite-system simulation, we ignore the decay of $J_{ij}^{\alpha \beta }$
with distance, so the system may include unphysically long-range
interactions and could simulate Hamiltonians in any dimensions. To retrieve $%
J_{ij}^{\alpha \alpha }$, for example, we start with a product state of all
spins, and perform time evolution using the entire Hamiltonian from Eq.~%
\eqref{Eq:HamGeneral}, interspersed with the $XY$-$8$ DD pulses on target spins $i$
and $j$. We would like to emphasize that specific state initialization for
the environment spins is not required as long as they are disentangled from
the target pair of qubits at the beginning. After $N_{c}$ cycles of the DD
sequence, the environment spins are traced out and measurements are made on
spins $i$ and $j$. In the simulation, we do not assume the pure unitary
evolution $U_{2\text{-spin}}$ as the remnant coupling to the environment
spins may entangle the two spins with the rest. However, any undesired
couplings are suppressed to the order of $O(J^{3}\tau ^{3})$ and we do
observe that the two-spin density matrix remains mostly pure ($\sim 99.9\%$)
for our chosen parameters.

As the tomography procedure involves measuring the output probability of a
certain state, each time point will be measured $N_{m}$ times, which gives
an estimate of the probability $p_{m}$ in this state. The measurement
uncertainty (standard deviation) will be $\sqrt{p_{m}(1-p_{m})/N_{m}}$
following the binomial distribution. As discussed above, to map out $%
c_{1},c_{2}$ and $c_{3}$, one needs to measure $P_{|+\text{I}\rangle
\rightarrow |00\rangle },P_{|+\text{I}\rangle \rightarrow |10\rangle }$ and $%
P_{|0\text{I}\rangle \rightarrow |++\rangle }$ for the target spins at
various time points and extract the corresponding oscillation frequencies.
Suppose $N_{t}$ different time points are measured for each set. The
oscillation frequencies can be found either by Fourier transform or by curve
fitting. In general, if data show numerous oscillation periods, Fourier
transform will be more robust and reliable \cite{Cole2005Identifying,
Cole2006Precision, Devitt2006Scheme}. In our case, however, the long time
observations will be undermined by the remnant coupling to the environment
spins and possible pulse error accumulation. Simple curving fitting with
fewer oscillation periods, therefore, appears to be a better solution. In
Fig.~\ref{Fig:Estimation}(a-c), we fit the data with the method of least
squares with $\tau J=0.01,N_{m}=100,N_{t}=50$ for spins $i=7$ and $j=9$ in a
$N=12$ spin system. The blue solid lines are the best-fit lines, and the red
dashed lines are the theoretical lines using the true coupling coefficients.
The longest time period requires $800$ pulses, which is well within the
current experimental technology without significant pulse error
accumulation. Table \ref{Table:Estimation} compares the true values and the
estimated ones of $J_{79}^{\alpha \alpha }$. Uncertainties in the estimation
stem from the curve fitting due to measurement uncertainties. Corresponding
results for $b_{6}^{\alpha }$ of spin $6$ are shown in Fig.~\ref%
{Fig:Estimation}(d-e) and Table \ref{Table:Estimation}. All estimated
parameters are accurate within a few percent.

To simulate real experiments, one also needs to include possible pulse
errors. One possible source of errors is the finite duration of each control
pulse, which limits the minimum cycle time. This is typically not the
dominant source of errors and can often be well-controlled \cite%
{Khodjasteh2007Performance, Viola2003Robust, Souza2012Robust,
Hodgson2010Towards, Uhrig2010Efficient}. In most experiments, the major
cause of errors is the deviation between the control pulses and the ideal $X$
or $Y$ pulses. These can either arise from the amplitude error where the
rotation angle differs from the ideal $\pi $-pulse or the rotation error
where the rotation axis deviates from the $x$ or $y$ axis. In typical
experiments, individual pulse errors may be controlled within a percent
level. In our simulation, we consider three different forms of pulse errors:
Systematic Amplitude pulse Error (SAE), Random Amplitude pulse Error (RAE)
and Random Rotation axis Error (RRE). See the caption of Table \ref%
{Table:Estimation} for the specific forms of the errors. Moderate systematic
errors can be self-compensated by the $XY$-$8$ DD sequence. Numerically, we
found that $5\%$ of SAE has negligible effect on the parameter estimation.
In addition, we also simulated the cases where each pulse experiences a $1\%$
RAE or RRE. Results are summarized in Table \ref{Table:Estimation}. The
average deviation from the true parameters are within $5\%$. Here, we would
like to point out a few features of our scheme that make it inherently
robust to errors. First of all, the estimation of the coupling strength $%
J_{mn}^{\alpha \beta }$ only entails frequency estimation, which could
endure large deviations of a few measurement points. In addition, the
single-parameter curve fitting scheme not only makes the estimation robust
but is also more convenient for experiment. Moreover, the retrieval of local
fields $b_{m}^{\alpha }$ is remarkably tolerant to pulse errors. Since no
pulse is directly applied to the target spin, any pulse errors on the
environment spins will only be propagated via the remnant DD coupling error,
which is suppressed to the order of $O(J^{3}\tau ^{3})$. We have numerically
tested that a $10\%$ pulse error of any kind would have negligible effects
on the estimation of $b_{m}^{\alpha }$. Alternative schemes to extract the
local fields are detailed and discussed in the Supplemental Material \cite%
{HamTom:Sup}. They are less tolerant to pulse errors, but may be easier to
implement in some experimental setups.

\textit{Discussion and outlook.---}We have thus numerically demonstrated
that the proposed scheme is robust to various sources of errors present in
real experiments. The measurement uncertainties can be lowered by increasing
$N_{m}$ and the pulse errors can be reduced by limiting the maximum number
of pulses needed. The optimal strategy involves a delicate balance between
experimental sophistication and error control. For example, by fixing $\tau J
$ and the total number of measurements for each set, $N_{m}\times N_{t}$,
one could devise an optimal estimation procedure. In addition, it is also
possible to eliminate the remnant DD coupling error to a higher order with
more elaborate pulse sequences such as the concatenated DD sequence \cite%
{Khodjasteh2005Fault, Khodjasteh2007Performance} and reduce pulse errors by
designing composite pulses or self-correcting sequences \cite%
{Souza2011Robust, Souza2012Robust, Levitt1986Composite, Ryan2010Robust}. The
scheme can also be extended straightforwardly to qudit systems of higher
spins or to bosonic or fermionic systems.

In conclusion, we have proposed a general scheme to achieve full Hamiltonian
tomography for generic interacting qubit systems with arbitrary long-range
couplings. The required number of measurements scales linearly with the
number of terms in the Hamiltonian, and the scheme is robust to typical
experimental errors or imperfections.

\begin{table}[t]
\caption{NPE: No Pulse Error; SAE: Systematic Amplitude pulse Error; RAE:
Random Amplitude pulse Error; RRE: Random Rotation axis Error; AD: Average
Deviation from true values. The last digit in bracket for each number
quantifies the estimation error bar due to measurement uncertainties, which is generated by the bootstrapping method. The percentage values in the brackets denote the amount of errors introduced
in each pulse. The errors are in the form of: SAE, $e^{i \frac{\protect\pi}{2%
} (1+\protect\epsilon)\protect\sigma^{\protect\nu}}$; RAE, $e^{i \frac{%
\protect\pi}{2} (1+\protect\delta)\protect\sigma^{\protect\nu}}$; RRE, $e^{i
\frac{\protect\pi}{2} (\protect\sigma^{\protect\nu} + \protect\alpha \protect%
\sigma^{x} + \protect\beta \protect\sigma^{y} + \protect\gamma \protect\sigma%
^{z} )}$; where $\protect\epsilon=5\%$, $\protect\delta$ is randomly chosen
from $(-1\%, 1\%)$, $(\protect\alpha,\protect\beta,\protect\gamma)$ is a
vector with a random direction but fixed magnitude at $1\%$, and $\protect\nu%
=x,y$ for the $X$ and $Y$ pulses respectively. }
\label{Table:Estimation}%
\begin{ruledtabular}
\begin{tabular}{cccccc}
& True & \multicolumn{4}{c}{Estimated Parameters}
\\ \cline{2-2} \cline{3-6}
 & -- &  NPE & SAE(5\%) & RAE(1\%) & RRE(1\%)
 \\ \hline
 $J_{79}^{xx}$ & $-0.378$  & $-0.369(3)$ & $-0.377(3)$  & $-0.379(4)$ & $-0.412(2)$
 \\
 $J_{79}^{yy}$ & 0.863  & 0.856(3)  & 0.846(3) & 0.867(4) & 0.836(2)
 \\
$J_{79}^{zz}$  & 0.679 &  0.669(5)   & 0.649(5) & 0.718(6) & 0.611(4)
\\
$b_{6}^{x}$ & 0.334  & 0.32(1) & 0.32(1) & 0.32(1) & 0.32(1)
\\
$b_{6}^{y}$  & 0.569 & 0.567(8)  & 0.567(8) & 0.567(8) & 0.568(8)
\\
$b_{6}^{z}$   & $-0.431$ & $-0.441(8)$ & $-0.443(8)$ & $-0.441(8)$ & $-0.441(8)$
\\ \hline
AD & -- & 2\% & 3\% & 3\% & 5\%
\\
\end{tabular}
\end{ruledtabular}
\end{table}

\begin{acknowledgments}
We would like to thank Z.-X. Gong for discussions. This work is supported by the IARPA MUSIQC program, the ARO and the AFOSR MURI program.
\end{acknowledgments}


%

\onecolumngrid

\newpage

\section{Supplemental Material: Hamiltonian tomography for quantum many-body systems with arbitrary couplings}

\begin{quote}
In this Supplemental Material, we provide more details on the dynamical decoupling scheme and error estimation, the retrieval of local field parameters and alternative schemes, and include further results taking into account of pulse errors.
\end{quote}

\section{Dynamical Decoupling}

The most general Hamiltonian with two-body qubit interactions can be written
as 
\begin{equation}
H=\sum_{\alpha ,\beta ,m<n}J_{mn}^{\alpha \beta }\sigma _{m}^{\alpha }\sigma
_{n}^{\beta }+\sum_{m,\alpha }b_{m}^{\alpha }\sigma _{m}^{\alpha }.
\end{equation}%
The energy unit of the Hamiltonian is taken to be $J$ such that $%
J_{mn}^{\alpha \beta }/J$ and $b_{m}^{\alpha }/J$ are bounded between $-1$
and $1$. The symmetric $XY$-$4$ dynamical decoupling (DD) sequence on both
spins $i$ and $j$ produces 
\begin{equation}
U_{1}=U_{0}^{1/2}\left( \sigma _{i}^{x}\sigma _{j}^{x}U_{0}\sigma
_{i}^{x}\sigma _{j}^{x}\right) \left( \sigma _{i}^{z}\sigma
_{j}^{z}U_{0}\sigma _{i}^{z}\sigma _{j}^{z}\right) \left( \sigma
_{i}^{y}\sigma _{j}^{y}U_{0}\sigma _{i}^{y}\sigma _{j}^{y}\right)
U_{0}^{1/2},
\end{equation}%
where $U_{0}=e^{-iH\tau }$ and $\tau $ is the time interval between
consecutive pulses. We can decompose $H$ into two parts. 
\begin{align}
H& =H_{0}+H_{1}, \\
H_{0}& =J_{ij}^{xx}\sigma _{i}^{x}\sigma _{j}^{x}+J_{ij}^{yy}\sigma
_{i}^{y}\sigma _{j}^{y}+J_{ij}^{zz}\sigma _{i}^{z}\sigma _{j}^{z}, \\
H_{1}& =J_{ij}^{xy}\sigma _{i}^{x}\sigma _{j}^{y}+J_{ij}^{xz}\sigma
_{i}^{x}\sigma _{j}^{z}+J_{ij}^{yx}\sigma _{i}^{y}\sigma
_{j}^{x}+J_{ij}^{yz}\sigma _{i}^{y}\sigma _{j}^{z}+J_{ij}^{zx}\sigma
_{i}^{z}\sigma _{j}^{x}+J_{ij}^{zy}\sigma _{i}^{z}\sigma _{j}^{y}  \notag \\
& +\sigma _{i}^{x}B_{i}^{x}+\sigma _{i}^{y}B_{i}^{y}+\sigma
_{i}^{z}B_{i}^{z}+\sigma _{j}^{x}B_{j}^{x}+\sigma _{j}^{y}B_{j}^{y}+\sigma
_{j}^{z}B_{j}^{z}+B,
\end{align}%
where $B_{i}^{\alpha }$ includes the local field on the $i$th spin and
interacting terms between the $i$th spin and all other spins other than the $%
j$th spin, i.e., $B_{i}^{\alpha }=b_{i}^{\alpha }+\sum_{\beta ,n\neq
i,j}J_{in}^{\alpha \beta }\sigma _{n}^{\beta }$. The bath term $B$ includes
all the environment operations, i.e., all operators that does not act on
spins $i$ and $j$. We define other Hamiltonian part as 
\begin{equation}
\sigma _{i}^{x}\sigma _{j}^{x}H\sigma _{i}^{x}\sigma
_{j}^{x}=H_{0}+H_{2},\quad \sigma _{i}^{y}\sigma _{j}^{y}H\sigma
_{i}^{y}\sigma _{j}^{y}=H_{0}+H_{3},\quad \sigma _{i}^{z}\sigma
_{j}^{z}H\sigma _{i}^{z}\sigma _{j}^{z}=H_{0}+H_{4},
\end{equation}%
where 
\begin{align}
H_{2}& =-J_{ij}^{xy}\sigma _{i}^{x}\sigma _{j}^{y}-J_{ij}^{xz}\sigma
_{i}^{x}\sigma _{j}^{z}-J_{ij}^{yx}\sigma _{i}^{y}\sigma
_{j}^{x}+J_{ij}^{yz}\sigma _{i}^{y}\sigma _{j}^{z}-J_{ij}^{zx}\sigma
_{i}^{z}\sigma _{j}^{x}+J_{ij}^{zy}\sigma _{i}^{z}\sigma _{j}^{y}  \notag \\
& +\sigma _{i}^{x}B_{i}^{x}-\sigma _{i}^{y}B_{i}^{y}-\sigma
_{i}^{z}B_{i}^{z}+\sigma _{j}^{x}B_{j}^{x}-\sigma _{j}^{y}B_{j}^{y}-\sigma
_{j}^{z}B_{j}^{z}+B, \\
H_{3}& =-J_{ij}^{xy}\sigma _{i}^{x}\sigma _{j}^{y}+J_{ij}^{xz}\sigma
_{i}^{x}\sigma _{j}^{z}-J_{ij}^{yx}\sigma _{i}^{y}\sigma
_{j}^{x}-J_{ij}^{yz}\sigma _{i}^{y}\sigma _{j}^{z}+J_{ij}^{zx}\sigma
_{i}^{z}\sigma _{j}^{x}-J_{ij}^{zy}\sigma _{i}^{z}\sigma _{j}^{y}  \notag \\
& -\sigma _{i}^{x}B_{i}^{x}+\sigma _{i}^{y}B_{i}^{y}-\sigma
_{i}^{z}B_{i}^{z}-\sigma _{j}^{x}B_{j}^{x}+\sigma _{j}^{y}B_{j}^{y}-\sigma
_{j}^{z}B_{j}^{z}+B, \\
H_{4}& =+J_{ij}^{xy}\sigma _{i}^{x}\sigma _{j}^{y}-J_{ij}^{xz}\sigma
_{i}^{x}\sigma _{j}^{z}+J_{ij}^{yx}\sigma _{i}^{y}\sigma
_{j}^{x}-J_{ij}^{yz}\sigma _{i}^{y}\sigma _{j}^{z}-J_{ij}^{zx}\sigma
_{i}^{z}\sigma _{j}^{x}-J_{ij}^{zy}\sigma _{i}^{z}\sigma _{j}^{y}  \notag \\
& -\sigma _{i}^{x}B_{i}^{x}-\sigma _{i}^{y}B_{i}^{y}+\sigma
_{i}^{z}B_{i}^{z}-\sigma _{j}^{x}B_{j}^{x}-\sigma _{j}^{y}B_{j}^{y}+\sigma
_{j}^{z}B_{j}^{z}+B.
\end{align}%
Basically, each term will either commute or anticommute with the operator $%
\sigma _{i}^{\alpha }\sigma _{j}^{\alpha }$. Those commuting with it will be
left invariant, and those anticommuting will have a flipped sign. $H_{0}$
and $B$ commute with each operator $\sigma _{i}^{\alpha }\sigma _{j}^{\alpha
}$, so they are left unchanged. Now we can see explicitly that $%
H_{1}+H_{2}+H_{3}+H_{4}=4B$, which is why the DD sequence effectively
decouples the two spins $i$ and $j$ with the rest of the spins. To estimate
the error, we combine the unitary evolution for a period and repeatedly make
use of the formula 
\begin{equation}
e^{\tau A}e^{\tau B}=e^{\tau A+\tau B+\frac{1}{2}\tau ^{2}[A,B]+O(\tau
^{3})}.
\end{equation}%
Ignoring $\tau ^{3}$ and higher-order terms, we find 
\begin{align}
U_{1}& =e^{-i\tau /2(H_{0}+H_{1})}e^{-i\tau (H_{0}+H_{2})}e^{-i\tau
(H_{0}+H_{4})}e^{-i\tau (H_{0}+H_{3})}e^{-i\tau /2(H_{0}+H_{1})}  \notag \\
& =e^{-i4\tau (H_{0}+B)+C},
\end{align}%
where the remnant coupling noise term is 
\begin{align}
C=& -\tfrac{1}{4}\tau ^{2}\left[ H_{0}+H_{1},H_{0}+H_{2}\right] -\tfrac{1}{2}%
\tau ^{2}\left[ \tfrac{3}{2}H_{0}+\tfrac{1}{2}H_{1}+H_{2},H_{0}+H_{4}\right] 
\notag \\
& -\tfrac{1}{2}\tau ^{2}\left[ \tfrac{5}{2}H_{0}+\tfrac{1}{2}%
H_{1}+H_{2}+H_{4},H_{0}+H_{3}\right] -\tfrac{1}{4}\tau ^{2}\left[ \tfrac{7}{2%
}H_{0}+\tfrac{1}{2}H_{1}+H_{2}+H_{3}+H_{4},H_{0}+H_{1}\right] +O(\tau ^{3}) 
\notag \\
=& -\tau ^{2}\left\{ [H_{0},H_{3}-H_{2}]+\tfrac{1}{2}[H_{2}-H_{3},H_{4}]+%
\tfrac{1}{2}[H_{2},H_{3}]\right\} +O(\tau ^{3}).
\end{align}%
In the error term $C$, the biggest contribution comes from terms like $%
[B,\sigma _{m}^{\alpha }B_{m}^{\alpha }]$. Our aim is to show that the error
does not scale with the system size $N$, i.e., $C=O(J^{2}\tau ^{2})$. Let us
consider one such term and write it out explicitly (suppressing the $\alpha
,\beta $ summation): 
\begin{equation}
\lbrack B,\sigma _{i}^{x}B_{i}^{x}]\sim \sigma _{i}^{x}\Bigg[\sum_{\substack{
m<n \\ m,n\neq i,j}}J_{mn}^{\alpha \beta }\sigma _{m}^{\alpha }\sigma
_{n}^{\beta }\;,\sum_{p\neq i,j}J_{ip}^{x\gamma }\sigma _{p}^{\gamma }\Bigg]%
\sim \sigma _{i}^{x}\sum_{\substack{ m<n \\ m,n\neq i,j}}J_{mn}^{\alpha
\beta }J_{im}^{x\gamma }\sigma _{m}^{\delta }\sigma _{n}^{\beta }.
\end{equation}%
Since $J_{mn}^{\alpha \beta }$ and $J_{im}^{x\gamma }$ are rapidly decaying
functions of the separation distance, for a fixed site $i$, $%
\sum_{m<n}J_{mn}^{\alpha \beta }J_{im}^{x\gamma }=O(J^{2})$. Note that this
differs from the scaling of the Hamiltonian, $H\sim \sum_{m<n}J_{mn}^{\alpha
\beta }=O(NJ)$. All the other terms in $C$ are either smaller or contribute
to the same order as the above term. Therefore, we have $C=O(J^{2}\tau ^{2})$%
. To be able to neglect the error terms, one needs to fulfill the condition $%
J\tau \ll 1$.

The above discussion is pertinent to the $XY$-$4$ pulse sequence. We can
cancel the second order contribution by using the $XY$-$8$ pulse sequence as 
$U_{2}=U_{1}U_{1}^{\text{R}}$, where $U_{1}^{\text{R}}$ is just the
time-reversed sequence of $U_{1}$. It can be readily seen that the error
terms $C$ and $C^{\text{R}}$ will cancel each other to the second order $%
O(\tau ^{2})$, since $C^{\text{R}}$ contains the same terms as in $C$ only
with the role of $H_{2}$ and $H_{3}$ interchanged. Therefore, the remnant
coupling error of the $XY$-$8$ pulse sequence is $O(J^{3}\tau ^{3})$ as
discussed in the main text.

\section{Local Field Retrieval}

In the main text, we proposed a scheme to retrieve the local fields $%
b_{i}^{\alpha }$ by shining the $XY$-$8$ pulse sequences on all the
environment spins. Here, we provide more details and outline alternative
schemes that may in some experimental setups be easier to implement. By
decoupling the environment spins with spin $i$ as illustrated in Fig.~1(c)
of the main text, we have the effective single-spin Hamiltonian $H_{1\text{%
-spin}}=b_{i}^{x}\sigma _{i}^{x}+b_{i}^{y}\sigma _{i}^{y}+b_{i}^{z}\sigma
_{i}^{z}$. The time evolution operator is 
\begin{equation}
U_{1\text{-spin}}=e^{-iH_{1\text{-spin}}T}=\left( 
\begin{array}{cc}
\cos (bT)-i\dfrac{b_{i}^{z}}{b}\sin (bT) & -\dfrac{ib_{i}^{x}+b_{i}^{y}}{b}%
\sin (bT) \\ 
\dfrac{b_{i}^{y}-ib_{i}^{x}}{b}\sin (bT) & \!\cos (bT)+i\dfrac{b_{i}^{z}}{b}%
\sin (bT)%
\end{array}%
\right) ,
\end{equation}%
where $b=\sqrt{(b_{i}^{x})^{2}+(b_{i}^{y})^{2}+(b_{i}^{z})^{2}}$ is the
magnitude of the Bloch vector. By measuring 
\begin{align}
P_{|0\rangle \rightarrow |0\rangle }& =1+\left[ \left( b_{i}^{z}/b\right)
^{2}-1\right] \sin ^{2}(bT) \\
P_{|+\rangle \rightarrow |+\rangle }& =1+\left[ \left( b_{i}^{x}/b\right)
^{2}-1\right] \sin ^{2}(bT)
\end{align}%
at various time points, we could determine $%
|b_{i}^{x}|,|b_{i}^{y}|,|b_{i}^{z}|$. To pin down the correct signs, one can
supplement the above two sets of measurements with another two measurement
points: 
\begin{align}
P_{|+\rangle \rightarrow |0\rangle }=\left\vert \langle 0|U_{1\text{-spin}%
}|+\rangle \right\vert ^{2}& =\dfrac{1}{2}\left( 1+\dfrac{2b_{i}^{x}b_{i}^{z}%
}{b^{2}}\sin ^{2}bT-\dfrac{b_{i}^{y}}{b}\sin 2bT\right)  \\
P_{|\text{I}\rangle \rightarrow |0\rangle }=\left\vert \langle 0|U_{1\text{%
-spin}}|\text{I}\rangle \right\vert ^{2}& =\dfrac{1}{2}\left( 1+\dfrac{%
2b_{i}^{y}b_{i}^{z}}{b^{2}}\sin ^{2}bT+\dfrac{b_{i}^{x}}{b}\sin 2bT\right) .
\end{align}%
Only one time point is needed to determine the signs. For example, one could
take measurements at $bT=\pi /4$ and use $P_{|+\rangle \rightarrow |0\rangle
}$ and $P_{|\text{I}\rangle \rightarrow |0\rangle }$ to pick out the correct
signs.

\begin{figure}[t]
\includegraphics[trim=0cm 0cm 0cm 0cm,
clip,width=0.6\textwidth]{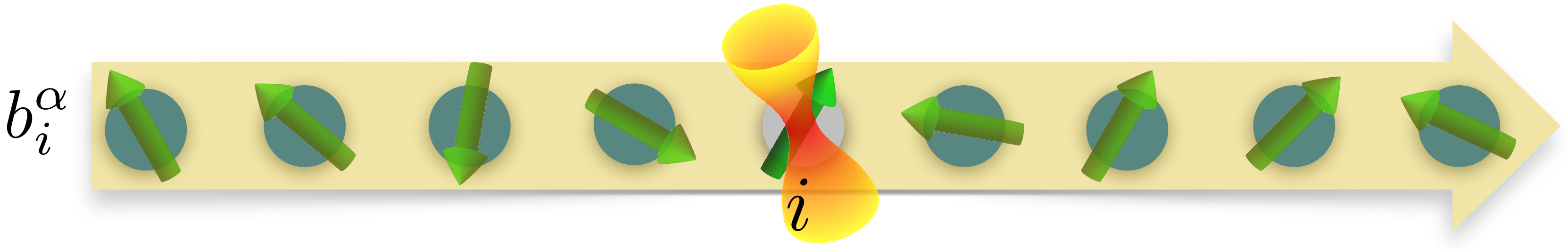}
\caption{Alternative scheme to map out the local fields $b_{i}^{\protect%
\alpha}$. A global pulse imposes the $XY$-$8$ pulse sequence on all spins
and a focused pulse is in addition applied to spin $i$ to cancel the DD
sequence on that single spin.}
\label{Fig:SingleSpin}
\end{figure}

The above procedure requires applying the DD sequences to all spins other
than the target spin. In some experimental setting, it may be easier to
apply a global DD sequence to all spins and add another individually
addressed beam on spin $i$ to cancel the DD sequence on that single spin.
See Fig.~\ref{Fig:SingleSpin} for illustration. For instance, one could
apply synchronized $X_{\text{All}}Y_{\text{All}}$-$8$ global pulses and in
addition $X_{i}Y_{i}$-$8$ focused pulses on spin $i$. In this way, spin $i$
effectively experiences no pulses at all time. The effective Hamiltonian
again reduces to the same $H_{1\text{-spin}}$ as above. However, this scheme
is not very robust to pulse errors. Any deviation from the ideal pulse will
be doubled on spin $i$ and accumulate. The pulse error will affect the
single-spin coherence and obscure $b_{i}^{\alpha }$ too. We have tested it
numerically that the pulse errors have to be controlled within $0.5\%$ for
the scheme to be feasible. So it can be used in some setups where pulse
errors are not an issue or the total number of pulses can be reduced. One
may also use this scheme and modify the sequence by designing composite
pulses or self-correcting sequences to reduce pulse errors.

\section{Pulse Errors}

\begin{figure}[h]
\includegraphics[trim=0cm 0cm 0cm 0cm, clip,width=\textwidth]{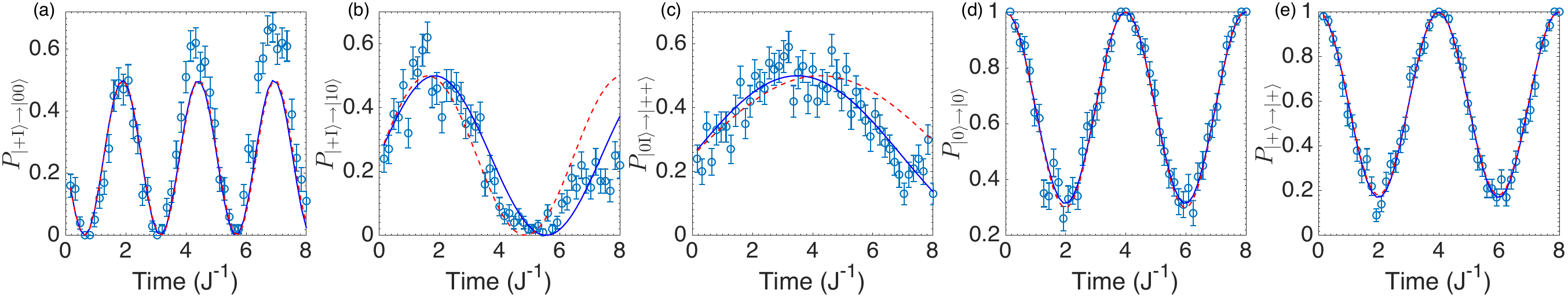}
\caption{ (color online). Numerical simulation and curving fitting results
with a Random Rotation axis Error (RRE) for each pulse. The RRE is of the form $%
e^{i \frac{\protect\pi}{2} (\protect\sigma^{\protect\nu} + \protect\alpha 
\protect\sigma^{x} + \protect\beta \protect\sigma^{y} + \protect\gamma 
\protect\sigma^{z} )}$ where $(\protect\alpha,\protect\beta,\protect\gamma)$
is a vector with a random direction but fixed magnitude at $1\%$. (a)-(c)
are used to retrieve $J_{79}^{xx}$, $J_{79}^{yy}$ and $J_{79}^{zz}$ between
spins 7 and 9. (d) and (e) are used to extract $b_{6}^{x}$, $b_{6}^{y}$ and $%
b_{6}^{z}$ for spin 6. The blue solid lines are the best-fit lines with the
simulated experimental data, and red dashed lines are the theoretical ones
generated by the true Hamiltonian parameters. Other parameters are the same
as in Fig.~2 of the main text.}
\label{Fig:EstimationRRE}
\end{figure}

In the main text, we discussed different types of pulse errors. In our
numerical simulation, we considered Systematic Amplitude pulse Error (SAE),
Random Amplitude pulse Error (RAE) and Random Rotation axis Error (RRE). The
fitting curves in Fig.~2 of the main text do not take into account of pulse
errors. Here, we include the figures (Fig.~\ref{Fig:EstimationRRE}) for the
case with a $1\%$ RRE. We can see, for example in Fig.~\ref%
{Fig:EstimationRRE}(a), that the frequency estimation is still very accurate
while some measurement points may have a notable mismatch. We may also
notice that the estimation of $b_{i}^{\alpha }$ is exceptionally robust to
pulse errors since no pulse is applied to spin $i$ in the scheme. Other
pulse errors have similar effects on the estimation of parameters. 

\end{document}